\documentclass[sigconf,10pt]{acmart}

\usepackage{caption}
\usepackage{subcaption}
\usepackage{cleveref}
\usepackage{enumitem}

\Crefname{figure}{Fig.}{Figs.}
\crefname{figure}{fig.}{figs.}

\crefformat{chapter}{\S#2#1#3}
\crefmultiformat{chapter}{\S\S#2#1#3}{ and~#2#1#3}{, #2#1#3}{, and~#2#1#3}

\crefformat{section}{\S#2#1#3}
\crefmultiformat{section}{\S\S#2#1#3}{ and~#2#1#3}{, #2#1#3}{, and~#2#1#3}

\def\withComments{1} 

\ifdefined\withComments
\newcommand{\authnote}[2]{{\bf [{\color{red} #1's Note:} {\color{blue} #2}]}}
\else
\newcommand{\authnote}[2]{}
\fi

\newcommand{\remove}[1]{}

\newcommand{\nparagraph}[1]{ \ \vspace{-0.2cm} \\  \noindent{\bf{#1}}}

\setcopyright{acmcopyright}

\acmDOI{10.1145/3538643.3539748}

\acmISBN{123-4567-24-567/08/06}

\acmConference[HotStorage'22]{14th ACM Workshop on Hot Topics in Storage and File Systems}{June 27-28, 2022}{virtual conference}
\acmYear{2022}
\copyrightyear{2022}

\begin{document}
  
\title{Rethinking Block Storage Encryption with Virtual Disks}

\author{Danny Harnik}
\affiliation{%
  \institution{IBM Research}
}

\author{Oded Naor}

\affiliation{%
  \institution{Technion\footnotemark[1]}
  \authornote{
Part of the work was done while Oded was interning at IBM Research.}
}

\author{Effi Ofer}
\affiliation{%
  \institution{IBM Research}
}

\author{Or Ozery}
\affiliation{%
  \institution{IBM Research}
}


\begin{abstract}
Disk encryption today uses standard encryption methods that are length preserving and do not require storing any additional information with an encrypted disk sector. 
This significantly simplifies disk encryption management as the disk mapping does not change with encryption. On the other hand, it forces the encryption to be deterministic when data is being overwritten and it disallows integrity mechanisms, thus lowering security guarantees. Moreover, because the most widely used standard encryption methods (like AES-XTS) work at small sub-blocks of no more than 32 bytes, deterministic overwrites form an even greater security risk. 
Overall, today's standard practice forfeits some security for ease of management and performance considerations. 
This shortcoming is further amplified in a virtual disk setting that supports versioning and snapshots so that overwritten data remains accessible.

In this work, we address these concerns and stipulate that especially with virtual disks, there is motivation and potential to improve security at the expense of a small performance overhead.
Specifically, adding per-sector metadata to a virtual disk allows running encryption with a random initialization vector (IV) as well as potentially adding integrity mechanisms. 
We explore how best to implement additional per-sector information in Ceph RBD, a popular open-source distributed block storage with client-side encryption. We implement and evaluate several approaches and show that 
one can run AES-XTS encryption with a random IV at a manageable overhead ranging from 1\%--22\%, depending on the
IO size.
\end{abstract}


%
%



\maketitle


\sloppy

\section{Introduction}

\nparagraph{Disk Encryption:}
{\em Data-at-rest disk encryption} is at the foundation of storage security and has been a central requirement for persistent storage over the years.
It requires that data is encrypted before being written to disk so that if the disk is stolen or illegally accessed, attackers would not be able to make sense of the data (as long as they do not hold the encryption key). 

When encrypting a disk one must take into account its structure and access patterns.
Disks are accessed at a sector granularity and as such, encryption is done at a sector-by-sector granularity.
Originally, disk sectors were 512 bytes each, and today they are typically 4096 bytes.
As such, disk encryption encrypts each sector separately.
Moreover, in a disk, each sector is addressed by the Logical Block Address (or LBA) and to simplify the integration of encryption, this mapping is kept intact when data is encrypted.
This implies that {\em the encryption of a sector should have the same length as the sector}.  

 Several issues arise when the length of the output cannot grow. Mainly: 
\begin{enumerate}[leftmargin=*]
    \item {\em Deterministic encryption:} To achieve {\em Semantically secure encryption}, the encryption must not be deterministic~\cite{GM84}. In particular, if a sector is overwritten, an adversary must not be able to determine whether the contents of the sectors have changed during the overwrite. Yet if the encryption is deterministic, this information is obvious since the same plaintext would yield the same ciphertext. The common mechanism to avoid determinism in encryption is to add a {\em nonce} as an input to the encryption. This nonce, usually called the Initialization Vector (IV), is a string of bits that is guaranteed not to repeat itself between instances of encryption, hence avoiding the determinism of the encryption function (note that the IV, unlike the encryption key, can be made public). 
    The IV is required in order to decrypt the data and hence must be stored alongside the encrypted data and read for the decryption process. The problem is that with standard disk encryption, there is no room left to store the IV alongside the encrypted sectors.
    \item {\em Authentication of encryption:} In length preserving encryption, changing a part of the cipher of a sector generates a new legal encryption pattern (of a different plaintext). This means that one cannot detect changes to the ciphertext, whether malicious or accidental. The common approach to handle authentication of encrypted data is to hold an additional Message Authentication Code (MAC) that can later be used to verify that the encrypted sector has not changed. In traditional disk encryption, there is no room to store a MAC associated with each sector.       
\end{enumerate}
In short, since the length of the output cannot grow, it is possible to identify which sections have changed between writes and also possible to revert a sector to an older version.

\nparagraph{Disk Encryption Today:}
Given the length preserving limitations and the lack of space for additional per-sector information, the cryptographic and security communities resorted to the following approach: 
\begin{itemize}[leftmargin=*]
    \item Use a unique data encryption key per disk. This key is used to encrypt all the sectors in the disk. 
    \item In order to avoid deterministic encryption across sectors, the {\bf sector number} or LBA is used as an additional per sector input to the encryption. The sector number is used together with the key to derive the actual IV used in each encryption block. Because the sector number is also known during reads, it can be used to correctly decrypt the data. 
    \item Devise encryption modes that will not ``break" if a nonce is repeated with different data. Namely, the data itself remains unknown, and the only information divulged is whether the underlying plaintext has changed or not. AES-XTS~\cite{Roagaway_XTS04, NIST_XTS} is the most commonly used method today that was designed to remain secure under repeating IVs. \footnote{ Historically, AES-CBC was the widely used encryption method, but it was replaced due to security attacks on this mode.}  
\end{itemize}

In reality, repeating the same IV even in AES-XTS is not ideal~\cite{XTS-blog}, as is explained below in \Cref{sec:XTS}.
However, it is a compromise that the community was willing to take with the lack of a better alternative. Since data written to different addresses uses different IVs, the only security concerns arise with overwrites to the same address. In that sense, full disk encryption with AES-XTS guarantees that if the disk is physically stolen, then no data in the disk is encrypted with the same IV since there is no record of overwrites of the data.\footnote{Note that in SSDs overwritten data typically remains on the disk until garbage collection kicks in. Thus, accessing the flash at a physical layer (rather than standard disk access) may reveal different data encrypted with the same IV (as pointed out in~\cite{ZJHZ+20}).}
It is only when an adversary eavesdrops to the write stream over a disk that it will encounter sectors encrypted with the same IV due to overwrites.     

\nparagraph{Virtual Disks:}
Virtual disks change the equation in two fundamental ways. The first is the snapshot capabilities which are an important feature of such disks. In the presence of snapshots, various versions of data written to the same sector are kept and persisted one alongside the other (this holds for every mechanism for snapshots or versioning above the disk layer, whether in virtual disks or not). This means that the guarantee of no repeating IV in a stolen physical disk no longer holds. With various versions of the data encrypted under the same IV one can, for example, manipulate the data to contain arbitrary combinations of data from various snapshots (creating the encryption of a data combination that was never actually written). 

The second consideration is that while for a physical disk it is very tempting to avoid adding a layer of virtual-to-physical mapping, for a virtual disk this is a non-issue. A virtual disk, by definition, already contains a virtual-to-physical mapping layer which we can piggyback on to augment the layout and incorporate additional per-sector information.

\nparagraph{Our Work:}
We study the possibilities of adding per-sector information in the context of encryption in a specific setting - Ceph block storage \cite{weil2006ceph} (also known as Ceph RBD).
Specifically, we modify the built-in client-side encryption in Ceph RBD to use a fresh random IV per each sector write. The IV is persisted to disk to be used during read operations.
We evaluate the tradeoffs of such a design, providing improved security at the cost of the overhead required to persist and read the random IVs. 
We show that for the best implementation option we test the performance overheads are no larger than a 22\% overhead on writes and 3\% on reads. 

\nparagraph{Structure:}
The rest of the paper is structured as follows: \Cref{sec:background} provides the necessary background and related work; \Cref{sec:implementation} details the implementation and results; and lastly, \Cref{sec:conclusion} concludes the paper and discusses future work.

\section{Background}
\label{sec:background}
\subsection{AES-XTS and its Shortcomings}\label{sec:XTS}
AES-XTS is the prevalent standard used for disk encryption to date.
Among others, it is available in Android~\cite{androidEncryption}, Apple's Filevault~\cite{fileVault}, Microsoft's BitLocker~\cite{bitLocker}, and Linux's DMCrypt~\cite{DMCrypto}.
It is a specific implementation of {\em tweakable encryption}~\cite{lRW11} - a block encryption mode that takes as input an additional parameter (the tweak) that can be public and adds variability to an otherwise deterministic encryption function.
AES-XTS is used in disk encryption by setting the tweak, also referred to as the IV (Initialization Vector), to be the sector number, also referred to as the LBA (Logical Block Address). Therefore, if the same data is written to different sectors they will result in totally different ciphertext as they will use different IVs.
A critical security property required of tweakable encryption is that if different plaintexts are encrypted with the {\em same IV}, still no information can be deduced about the encrypted data. AES-XTS only achieves this to a certain extent.   

In an ideal block cipher, even if it is deterministic, changing a single bit in the plaintext of a sector will result in an entirely different, (random-looking) ciphertext sector. However, AES-XTS falls short of this (as do many other AES-based encryption modes). In AES-XTS, changing a single bit in the sector (without changing the key or IV) will yield the expected change only to the sub-block in the cipher to which this bit belongs.
The sub-blocks are the same size as the encryption key -  either 32 bytes (AES-256) or 16 bytes (AES-128) and stem from the way AES-XTS is built on top of the AES primitive - a building block that works on small 16/32 byte blocks. Such ciphers are referred to as narrow-block encryption.
This means that during an overwrite of a sector (using the same LBA and thus the same IV), an adversary can detect exactly which of the sub-blocks has changed and which have remained the same. 
Moreover, one can manipulate ciphertexts at a sub-block level. For example, given two versions of ciphertexts written to the same LBA, one can generate a new ciphertext of this sector that combines sub-blocks from both versions. The resulting ciphertext is legal and the manipulation cannot be detected. So encrypting different plaintexts with the same IV provides very good security guarantees at a granularity of a single sub-block, yet leaks some information about the relation between the plaintexts at a sector granularity. 
Still, due to the practicality of AES-XTS for disk encryption, it is widely used.  

Similar shortcomings also exist in other popular block cipher modes.
For example, in AES-CBC~\cite{ehrsam1978message} one can detect the first sub-block in which a bit has changed. Note that other methods like AES-GCM~\cite{mcgrew2004galois,mcgrew2004security} are completely insecure if the same key and IV are used and may leak information about the plaintext. Hence such modes can only work with a true nonce as an IV (one that never repeats). 

\subsection{Possible Mitigations}
The approach that we take to remedy the security shortcomings is to use a {\bf random IV} rather than use the LBA. If the random IV is chosen from a large enough range, the probability of ever repeating an IV is negligible.
This in essence removes the determinism of encryption for overwriting a sector and an adversary would not be able to detect if the underlying plaintext has changed at all. 
However, as described in the introduction, this requires writing the per-sector IV to the disk so that the sector could be decrypted during reads. Note that one should also include the sector number as part of the IV in order to avoid replay attacks where data encrypted at one LBA is replayed at another LBA.\footnote{In a system with snapshots one can also integrate the snapshot number into the IV to avoid cross snapshot replay attacks.}

Using an authentication code (MAC) on the ciphertext can prevent the various manipulation attacks described above, but also requires additional space. Also, using authentication alone still exposes which parts of the plaintext have changed during an overwrite.  

Another approach is using {\em wide-block} encryption~\cite{IEEEwideblock}, an encryption method in which every bit of the plaintext of a sector will influence the entire ciphertext of the sector (as opposed to methods like XTS which are narrow-block ciphers). 
This holds even if it is built on top of a building block like AES which works on a much smaller size than the sector. Wide-block encryption has been standardized~\cite{IEEEwideblock}, with two certified methods - XCB-AES\cite{MF07XCB} and EME2-AES~\cite{Halevi04EME, IEEEwideblock}. Yet it has not been widely adopted mainly due to lower performance, as well as implementation and patenting considerations. 
Using a wide-block cipher still carries the limitations of a deterministic cipher (an exact overwrite is easily identified), yet limits the attack granularity to that of a full sector. 

\subsection{Related Work}\label{sec:related}
The security concerns about the commonly used methods for disk encryption have been raised and studied by the cryptographic community. This was the main motivation for studies on wide-block encryption and their standardization effort~\cite{IEEEwideblock}.

Bro{\v{z}} et al.~\cite{BPM18} studied adding additional per-sector metadata as part of the dm-crypt encryption framework in the Linux kernel. They do this by using an additional device-mapper called dm-integrity that can be used for storing authentication information, or in the encryption case also a random IV.
To ensure consistency between the data sector and its metadata, they resort to using a journal which is shown to reduce the throughput by nearly one-half. 
Zhang et al.\ ~\cite{ZJHZ+20} integrate the AES-XTS encryption with the Flash Translation Layer (FTL) of an SSD and use the number of overwrites a sector has as a seed for its IV (hence ensuring that each overwrite gets a unique IV). This approach works well for {\em storage-side} encryption, which means that the data exists in the clear {\em at the storage} before being encrypted. 
Our work targets encryption at the client-side of a distributed storage system, ensuring that the data is always encrypted outside of the client, and attempts to piggyback the indirection layer of the distributed storage system.   

Note that some storage protocols like NVMe (starting from version 1.2)~\cite{nvme} include the option for a per-sector metadata support. However, this is not widely implemented in existing SSDs and existing implementations typically only allow for 8 bytes of metadata per sector, which is too short for our use-case.  

\begin{figure}[t]
\centering
\includegraphics[width=0.3\textwidth]{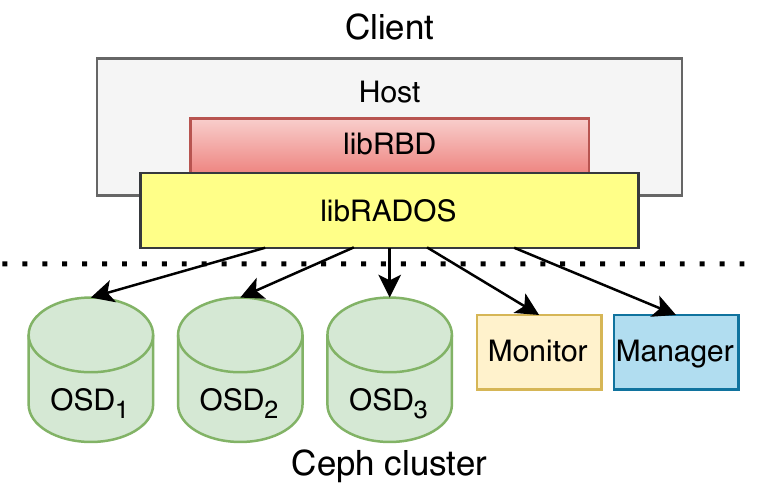}
\vspace{-1em}
\caption{
\label{fig:cephDiagram}%
Ceph's architecture}
\vspace{-1em}
\end{figure}

\subsection{Ceph RBD Encryption}
Ceph~\cite{weil2006ceph} is an open-source distributed storage platform that provides support for object storage, block storage, and file storage.
In this work, we focus on the block-storage of Ceph called RBD (Rados Block Device). 

The general architecture of a Ceph RBD deployment is depicted in \Cref{fig:cephDiagram}.
The Ceph cluster is made out of OSD nodes (\emph{Object Storage Devices}) that actually store the data and its replicas,  \emph{monitors} that maintain maps of the cluster state and perform access control, and \emph{managers} that provide additional metrics  and interfaces.

In its standard deployment, Ceph RBD employs a client-side driver called libRBD at each host accessing the storage.
For every virtual disk, libRBD maps each LBA to a specific OSD node by breaking the LBA space into objects (typically 4MB in size) and computing a placement algorithm for objects. The libRBD library distributes each IO to its corresponding OSD via a proprietary protocol called RADOS. The RADOS protocol supports several high-level functions such as snapshots. It also has supports transactions in which writes of several small IOs are guaranteed to be written atomically.
This proved very useful in ensuring consistency between written data and per-sector metadata.  

RBD supports client-side encryption~\cite{cephEncryption}, allowing for data to never leave the host in the clear. 
The encryption follows the LUKS standard for encrypted disks 
in which the default encryption is AES-XTS.\footnote{Note that LUKS has 2 versions, LUKS1~\cite{luks1} and LUKS2~\cite{luks2}.
In LUKS2, the default sector size is 4KB per block whereas in LUKS1 it is limited to 512 bytes only which makes adding per-sector information far more costly.
In this work, we only consider 4KB sectors.} 

\section{Implementation and Evaluation}
\label{sec:implementation}
\subsection{Design Choices for Storing Per-Sector Information}

\begin{figure}[t]
\centering
\begin{subfigure}[b]{0.4\textwidth}
\centering
\includegraphics[width=\textwidth]{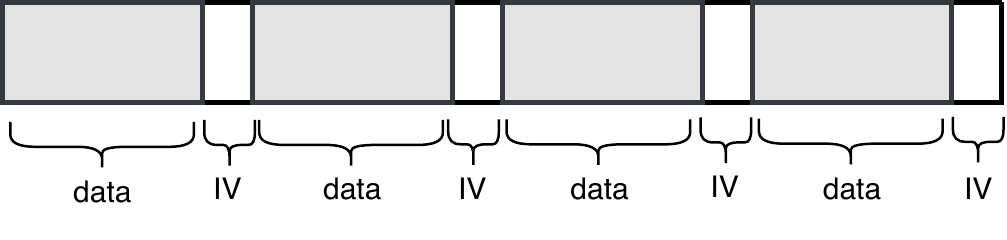}
\vspace{-2em}
\caption{Unaligned: each IV is stored at the end of its block.}
\label{fig:storageOptions:a}
\vspace{0.5em}
\end{subfigure}
\begin{subfigure}[b]{0.4\textwidth}
\centering
\includegraphics[width=\textwidth]{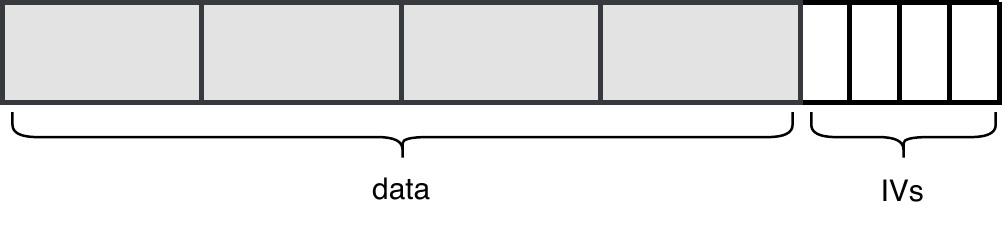}
\vspace{-2em}
\caption{Object end: All IVs stored at the end of the entire object.}
\label{fig:storageOptions:b}
\vspace{0.5em}
\end{subfigure}
\begin{subfigure}[b]{0.4\textwidth}
\centering
\includegraphics[width=\textwidth]{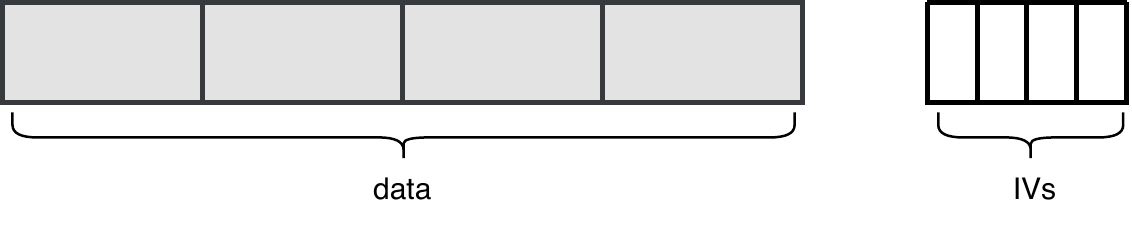}
\vspace{-2em}
\caption{OMAP: IVs stored at an external key-value DB.}
\label{fig:storageOptions:c}
\end{subfigure}
\vspace{-1em}
\caption{
\label{fig:storageOptions}%
Storage options for IVs}
\vspace{-1em}
\end{figure}

\begin{figure*}[t!]
\centering
\begin{subfigure}[b]{0.49\textwidth}
\centering
\includegraphics[width=\textwidth]{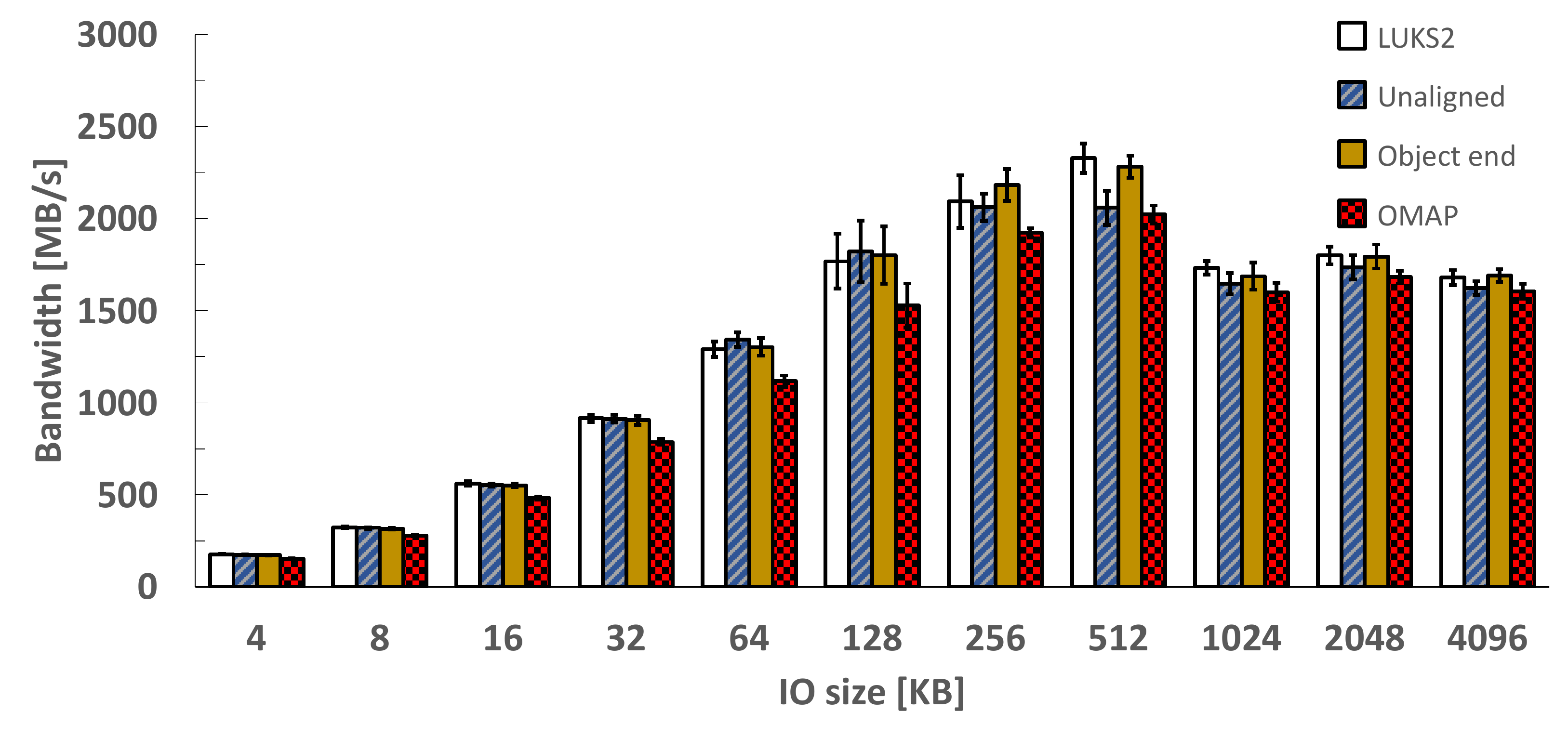}
\caption{Read bandwidth}
\end{subfigure}
\hfill
\begin{subfigure}[b]{0.49\textwidth}
\centering
\includegraphics[width=\textwidth]{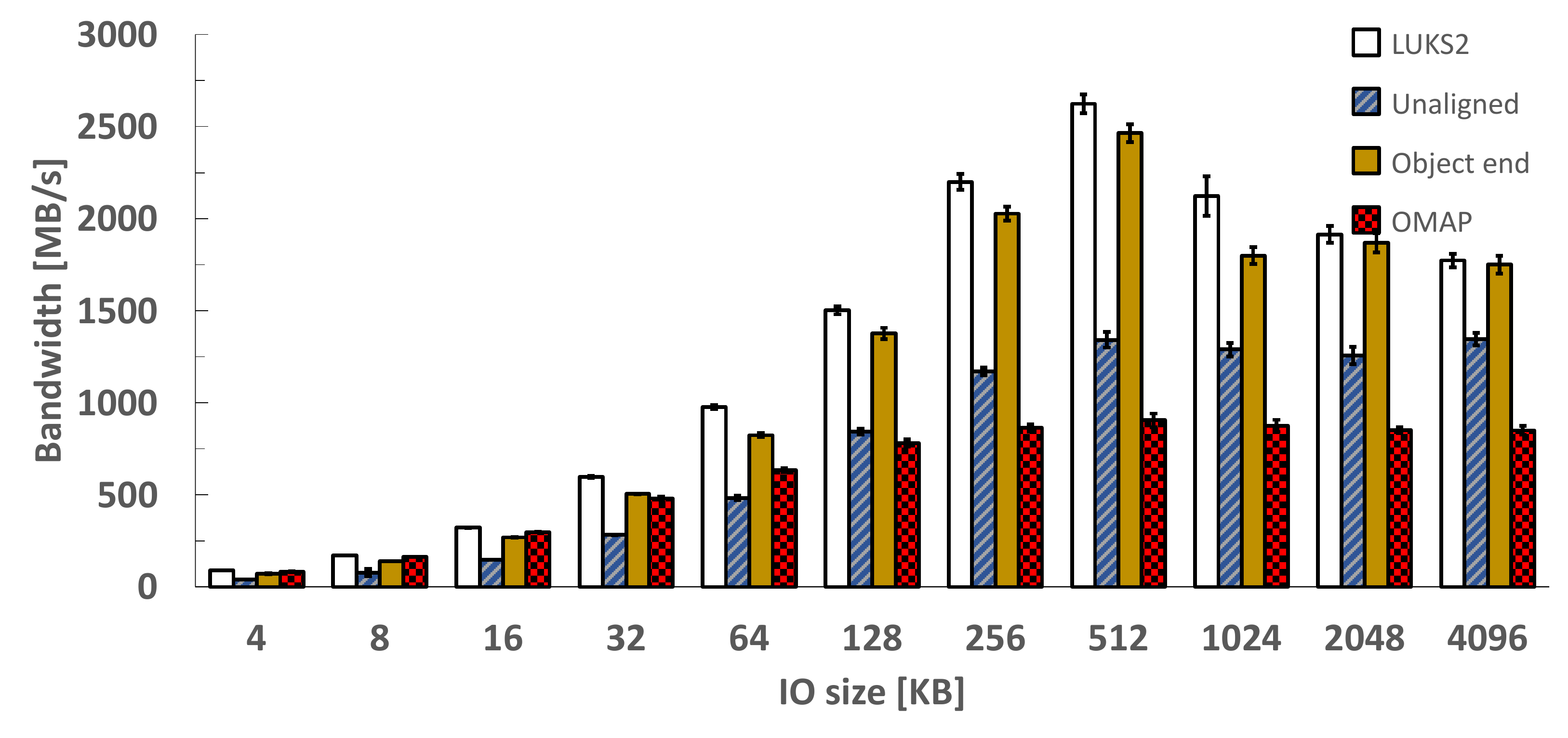}
\caption{Write bandwidth}

\end{subfigure}
\vspace{-0.5em}
\caption{
\label{fig:perfResults}%
Performance results for random read and write workloads} 
\end{figure*}

\begin{figure}[t]
\centering
\includegraphics[width=0.45\textwidth]{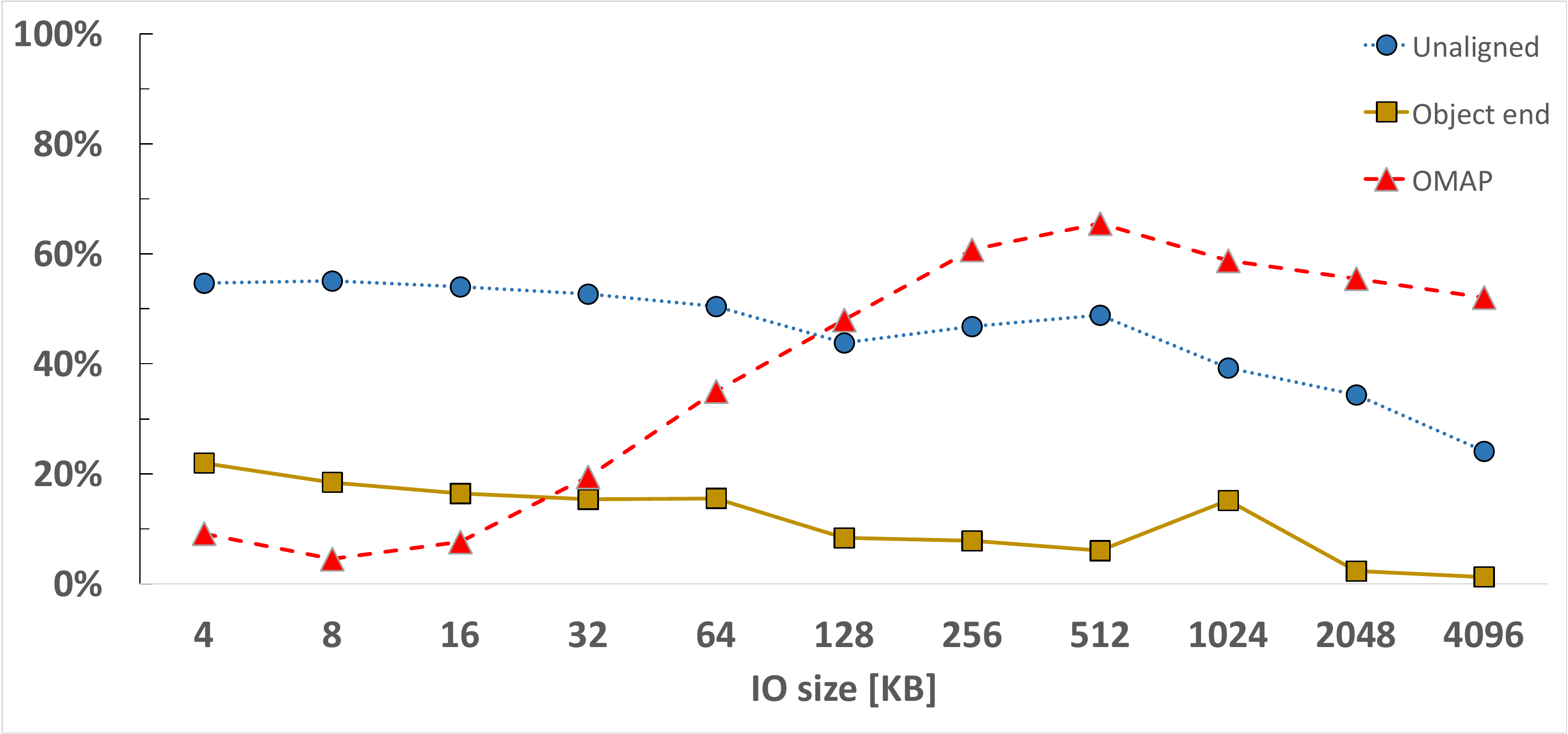}
\vspace{-0.5em}
\caption{
\label{fig:writeNormalized}%
Write performance overhead
}
\vspace{-2em}
\end{figure}
We explore how to integrate support for additional per-sector information. We focus on the use-case of using a random IV with AES-XTS encryption, but this can be used also for storing integrity information, or using an alternative cipher like AES-GCM. We leave evaluation of these options to future research.    

We implement three alternatives on where to store the IV with Ceph RBD, which are illustrated in \Cref{fig:storageOptions}.

The first alternative is the na\"ive approach of storing each IV after its matching block in an \emph{unaligned contiguous} manner.
Thus, each access is contiguous to the data and its matching IV, but since the IV size is less than a disk sector size, almost all of the data is unaligned to the disk sectors.

The second alternative keeps the sector alignment by packing several IVs together. The Ceph architecture 
of breaking data into objects lends itself nicely to this approach. For each object 4MB in size, we store all the IVs of the object after the encrypted data, i.e., at the \emph{object end}.
In this manner, we keep the division into objects as before, the address of an LBA within an object also remains the same, and the IVs are batched together at the object end. 

Lastly, we consider using a separate key-value database to store the IVs.
In Ceph, each object has a matching database to store additional metadata.
This database, which is named OMAP, is implemented using RocksDB~\cite{rocksDB} and can be used to store the IVs.
OMAP supports accessing multiple values based on a range of integer keys with a single operation.
Thus, if the key used to store each IV is the offset of the block in the object, then a contiguous read or write of the data can also be accessed with a single operation on the key-value database.

In all of the above implementations we use the support in the Ceph RADOS protocol for atomically writing multiple IOs to ensure data and IV consistency.   

\vspace{-0.3cm}
\subsection{Test Environment}\label{sec:env}

We implement the three alternatives and test their performance on a 3 node Ceph cluster
running Ceph version 16.2.4. We use Ceph's default configuration of 3-way replication, an object size of 4MB, and an encryption block size of 4KB. We only modify the code on the client nodes that run our modified libRBD code.
The OSD nodes are Intel Xeon E5-2650 v4 CPUs. Each node has 9 Intel NVMe disks, of 1.8TB each, and 128GB of memory by 8 DDR4-2400 CL17.
The OS is Red Hat Enterprise Linux version 8.4 (Ootpa).
The client nodes have a similar CPU and OS version, but with 384 GB memory by 12 DDR4-2400 CL17 of 32GB each.
The links between all nodes are 100 Gb/s, and when running iperf between the nodes we measure a bandwidth of around 13 Gb/s.

\vspace{-0.3cm}
\subsection{Results}\label{sec:results}
To measure the throughput performance, we use fio~\cite{fio} which has native support for Ceph RBD.
We use fio version 3.1, and deploy a single client random read and write workloads, with 32 maximum parallel accesses on a full Ceph image of 64GB. There are tests for IO sizes ranging from 4KB to 4MB and each test is repeated 10 times.
Sequential IO tests are not presented, but give similar results to random IO with large sizes. 

\Cref{fig:perfResults} presents a comparison of the three approaches to the baseline which is Ceph's LUKS2 implementation~\cite{cephEncryption} with deterministic LBA based IVs that are not stored.
Of the three random IV implementation options, the object end gives the best results for both reads and writes.

For the read workloads, all three approaches perform nicely, likely due to the backend's ability to do the IV reads in parallel to the data IO. The OMAP version fares slightly worse, due to the overhead of accessing the DB. The object end approach closely mirrors the baseline where the biggest difference we measure is $3\%$.

For the write workloads, there is a significant difference between the three options.
\Cref{fig:writeNormalized} presents the performance degradation of each method compared to the LUKS2 baseline, i.e., lower is better.
For the small block sizes, the OMAP solution gives the best performance, but this briefly changes as the IO size increases and the DB fails to provide high performance. 
The object end option performs better for almost all IO sizes, resulting in 1\%--22\% performance loss, depending on the IO size.

As the IO size grows, the theoretical overheads of unaligned and object end, measured as the number of sectors that need to be read or written to disk, decrease. For example, in a 4KB write/read, a minimum of two physical disk sectors need to be accessed (one for the data and one for the IV) versus one in the baseline. Whereas a 32KB IO typically requires 9 sectors to be accessed versus 8 in the baseline. Indeed, as the IO size grows we see an improvement in the measured performance overhead (except for an unexplained degradation at 1024KB writes). In the OMAP solution, this calculation does not work and hence the overhead grows significantly with the IO size. We suspect that the unaligned solution performs worse due to unaligned operations that trigger costly read-modify-write operations (these could potentially be further optimized with intelligent buffer writes).

\section{Conclusions and Looking Forward}
\label{sec:conclusion}
The purpose of this paper is to raise awareness in the storage community to the security concessions that we make in order to accommodate simpler and more efficient encryption techniques.
We further wish to demonstrate that better security can be achieved by adding the proper support in the storage layer and explore the performance tradeoff associated with this.
We hope to further understand the performance results that we observed and explore how much they apply to different Ceph configurations and different hardware or scale. We also ask how this design can be generalized to other systems.

We point out that working at the virtual mapping layer of the storage system creates opportunities for more efficient implementation than doing the mapping as an additional layer (like the implementation in dm-crypt~\cite{BPM18}). We expect that similar designs can be achieved in other architectures for virtual disk other than Ceph.

In the long term, we believe that block storage systems would benefit by natively supporting per-sector metadata as part of their initial design. Could a change in the standard block storage APIs to include per-sector metadata be advisable or beneficial? This could allow simple extensions from layers above the block storage (like dm-crypt). Furthermore, we wonder what additional usecases could benefit from per-sector metadata beyond security and integrity? 

\begin{acks}
Oded Naor is grateful to the Azrieli Foundation for the award of an Azrieli Fellowship, and to the Technion Hiroshi Fujiwara Cyber-Security Research Center for providing a research grant.
We thank Jonas Pfefferle, Nikolas Ioannou, Andreas D\"{o}ring and Sangeev Gupta for their help with the evaluation environments. We also thank Jason Dillaman for enlightening design discussions and his insights on Ceph architecture.
\end{acks}

\bibliographystyle{ACM-Reference-Format}
\bibliography{references}
\end{document}